\documentclass[12pt]{article}
\usepackage{amssymb,amsmath,epsfig}
\allowdisplaybreaks

\begin{document}

\title{\bf Dynamical Instability of Cylindrical Symmetric Collapsing Star in
Generalized Teleparallel Gravity}
\author{\bf {Abdul Jawad}$^{1}$
\thanks{jawadab181@yahoo.com;~~abduljawad@ciitlahore.edu.pk},
\bf{Davood Momeni}$^{2}$\thanks{d.momeni@yahoo.com},
\bf{Shamaila Rani}$^{1}$\thanks{drshamailarani@ciitlahore.edu.pk}\\
and
\bf{Ratbay Myrzakulov}$^{2}$\thanks{rmyrzakulov@gmail.com},\\
$^{1}$Department of Mathematics, COMSATS Institute of\\ Information
Technology, Lahore-54000, Pakistan.\\$^{2}$Eurasian International
Center for Theoretical Physics\\ and Department of General \&
Theoretical Physics,\\ Eurasian National University, Astana 010008,
Kazakhstan}

\date{}
\maketitle
\begin{abstract}
This paper is devoted to analyze the dynamical instability of a
self-gravitating object undergoes to collapse process. We take the
framework of generalized teleparallel gravity with cylindrical
symmetric gravitating object. The matter distribution is represented
by locally anisotropic energy-momentum tensor. We develop basic
equations such as dynamical equations along with matching conditions
and Harrison-Wheeler equation of state. By applying linear
perturbation strategy, we construct collapse equation which is used
to accomplish the instability ranges in Newtonian and post-Newtonian
regimes. We find these ranges for isotropic pressure as well as
reduce the results in general relativity. The unstable behavior
depends on matter, metric, mass and torsion based terms.
\end{abstract}

\section{Introduction}

The generalized teleprallel theory of gravity ($f(T)$ where $T$ is
torsion scalar) is one of the modified theories which occupied a
vast area of research in modern cosmology and Astrophysics. General
relativity is identified as the cornerstone of modern cosmology.
Einstein introduced cosmological constant, a simplest model which is
marked by cosmological observations on the basis of general
relativity (GR). This theory based on curvature via Levi-Civita
connection. This constant happened to inherit some flaws later like
fine-tuning and cosmic coincidence problems. Later on, Einstein
proposed a theory which is equivalent to GR established through
parallel transported of veirbein field, named as teleparallel
gravity. In this theory, the torsion scalar laid down the
gravitational field taking into account Weitzenb\"{o}ck connection.
The modified form of GR shapes $f(R)$ gravity while $f(T)$ is the
modification of teleparallel gravity by modifying curvature and
torsion scalars upto higher order terms.

Firstly, the $f(T)$ theory of gravity is proposed under Born-Infeld
strategy which helped to solve particle horizon problem and found
singularity-free solution (Ferraro and Fiorini 2008). Later, many
phenomena which includes expanding universe with acceleration
through different cosmological parameters, planes, perturbations,
cosmographic techniques, energy conditions, spherical symmetric
solutions via solar system constraints, Birkhoff's theorem,
connection between different regions of the universe through static
and wormhole solutions, viability of thermodynamical laws, unstable
behavior of spherical symmetric collapsing stars applying different
dynamical conditions, reconstruction scenario via dynamical dark
energy models, (Jamil et al. 2012;Jamil, Momeni \& Myrzakulov 2012,
2013, Houndjo, Momeni \& Myrzakulov 2012; Wang 2011a,b; Daouda,
Rodrigues \& Houndjo 2011, 2012; Gonzlez, Saridakis and Vsquez 2012;
Bamba, Nojiri \& Odintsov 2014; Sharif and Rani 2013, 2014a,b; Jawad
and Rani 2015) etc. The $f(T)$ theory of gravity is non local
Lorentz invariant theory. To resolve this problem, a lot of work has
been done in this direction (Li, Sotiriou \& Barrow 2011; Li, Miao
\& Miao 2011). Nashed (2015) proposed general tetrad field by
regularization of $f(T)$ field equations which has two tetrad
matrices. This regularized process with general tetrad field removed
the effect of local Lorentz invariance.

The dynamics of self-gravitating object during collapse process has
been analyzed in GR (Skripkin 1960; Chandrasekhar 1964; Herrera,
Santos \& Le Denmat 1989; Chan, Herrera \& Santos 1993; Herrera \&
Santos 1995; Herrera, Le Denmat and Santos \& 2012) and modified
theories of gravity such as $f(R), ~f(R,T)$, Brans-Dicke, Einstein
Gauss-Bonnet gravity in spherical as well as cylindrical symmetry
(Sharif \& Manzoor 2014,2015; Kausar, Noureen \& Shahzad 2015;
Kausar 2013) while in $f(T)$ gravity for spherical symmetry (Sharif
\& Rani 2014, 2015; Jawad \& Rani 2015). The disturbance in
hydrostatic equilibrium of a stellar object leads to the collapse
process. The instability occurs when the weight of the outer region
surrounded the object overcome very quickly the pressure inside the
object and gravitational force consequently pushes the matter
towards center of the object initiating collapse. In order to study
the unstable behavior of self-gravitating collapsing objects, the
adiabatic index which is a stiffness parameter is used. For
spherically symmetric matter configuration of collapsing star, this
index gives numerical range of the instability as less than 4/3 in
GR. However, in modified theories of gravity, there appear effective
terms results from modification of theory. Now-a-days, a lot of work
is done on jeans instability too (Roshan \& Abbass 2014, 2015).

In $f(R)$ gravity, Sharif \& Kausar (2011) and Kausar \& Noureen
(2014) analyzed the dynamical instability ranges for Newtonian and
post-Newtonian regimes of a spherical symmetric collapsing star with
and without charge. These ranges depend on matter, metric, mass and
curvature based terms. For cylindrical symmetry, Kausar (2013)
studied the effects of CDTT model which has inverse curvature term
on the unstable behavior. The asymptomatic behavior is also obtained
for both regimes. Sharif \& Manzoor (2014) examined the instability
of cylindrically symmetric collapsing object in the frame work of
Brans-Dicke gravity. They concluded that adiabatic index remains
less than one for the unstable behavior while greater than one in a
special case. In $f(T)$ gravity, Sharif \& Rani (2014, 2015)
analyzed the dynamics of self-gravitating object with spherical
symmetry via expansion and expansion free cases. Recently, Jawad \&
Rani (2015) examined the instability ranges taking into account
shear-free condition for Newtonian and post-Newtonian regimes.

The collapse process happens when stability of matter disturbed and
at long last experiences collapse which leads to different
structures. We have taken the self-gravitating object as
cylindrically symmetric collapsing star. The dynamical instability
analysis is mostly done for spherically symmetric object which
includes galactic halos, globular clusters, etc. However, the
non-spherical objects such that cylindrical symmetry and plates came
into being by post-shocked clouds on the verge of gravitational
collapse at stellar scales as well as galaxy formations. The
cylindrical symmetry is associated with the problem of fragmentation
of pre-stellar clouds. Specifically, final fate of collapse of a
non-spherical cloud coming from numerical relativity and certain
analytical solutions in cylindrical symmetry provide some new
examples about gravitational collapse.

In this connection, we extend our work on dynamical instability to
the cylindrically symmetric collapsing stars in $f(T)$ gravity. We
analyze the dynamical instability ranges in Newtonian and
post-Newtonian regimes. The paper is organized as follows: in next
section, we provide the basics of generalized teleparralel gravity.
Section 3 is devoted to the construction of some basic equations
such as field, dynamical and matching equations. In section 4, we
develop collapse equation for cylindrically symmetric collapsing
star. The instability ranges for anisotropic as well as isotropic
fluid for Newtonian and post-Newtonian regimes are examined in
section 5. Last section provides results of the paper.

\section{Generalized Teleparallel Gravity}

In this section, we provide the basics and formulation of
generalized teleparallel gravity. This gravity is defined through
the action (Ferraro \& Fiorini 2008; Linder 2010a,b; Bamba, Geng \&
Lee 2010; Wu \& Yu 2010a,b, 2011; Bamba et al. 2011; de la
Cruz-Dombriz, Dunsby \& Saez-Gomez 2014)
\begin{equation}\label{a}
\mathcal{I}=\frac{1}{\kappa^2 }\int d^4x (\mathcal{L}_m+f(T))h,
\end{equation}
where $f(T)$ is an arbitrary differentiable function, $T$ denotes
torsion scalar, $\mathcal{L}_m$ represents the matter density and
$\kappa^2$ denotes the coupling constant. The term
$h=\textmd{det}({h^a}_{\beta})$ is the determinant of vierbein (or
tetrad) field, ${h^a}_{\beta}$. This field has a basic and central
part in construction of this torsion based gravity. This is an
orthonormal set of vector fields related with metric tensor by the
relation
$g_{_{\beta\alpha}}=\eta_{ab}{h^a}_{\beta}{h^b}_{\alpha},~\eta_{ab}=\textmd{diag}(1,-1,-1,-1)$
is the Minkowski space. It is noted here that the indices
($a,b,...$) are the coordinates of tangent space while
($\alpha,\beta,...$) expresses the coordinate indices on manifold
and all these indices run from $0,1,2,3$. The parallel transported
of vierbein field by the significant component Weitzenb\"{o}ck
connection in Weitzenb\"{o}ck spacetime leads the basic construction
of teleparallel as well as generalized teleparallel gravity where
${\widetilde{\Gamma}^\beta}_{~\alpha\gamma}={h_a}^{\beta}\partial_\gamma{h^a}_\alpha$
is the Weitzenb\"{o}ck connection. We obtain torsion tensor by the
antisymmetric part of this connection as follows
\begin{equation}\label{1.2.10}
{T^\beta}_{\alpha\gamma}={\widetilde{\Gamma}^\beta}_{~\gamma\alpha}-
{\widetilde{\Gamma}^\beta}_{~\alpha\gamma}
={h_a}^{\beta}(\partial_{\gamma}{h^a}_{\alpha}-\partial_{\alpha}{h^a}_{\gamma}),
\end{equation}
which is antisymmetric in its lower indices, i.e.,
${T^\beta}_{\alpha\gamma}=-{T^\beta}_{\gamma\alpha}$. It is noted
that the parallel transported of tetrad field vanishes rapidly the
curvature of the Weitzenb\"{o}ck connection.

The torsion scalar takes the form
\begin{equation}\label{1.3.5}
T={S_\beta}^{\alpha\gamma}{T^\beta}_{\alpha\gamma}.
\end{equation}
where
\begin{equation}\label{1.3.6}
{S_\beta}^{\alpha\gamma}=\frac{1}{2}[\delta^{\alpha}_{\beta}{T^{\mu\gamma}}_{\mu}-
\delta^{\gamma}_{\beta}{T^{\mu\alpha}}_{\mu}-\frac{1}{2}(T^{\alpha\gamma}_{~~\beta}
-T^{\gamma\alpha}_{~~\beta}-T_{\beta}^{~\alpha\gamma})].
\end{equation}
Varying of action of $f(T)$ gravity w.r.t vierbein field, we obtain
the field equations as
\begin{equation}\label{1.3.9}
{h_a}^{\beta}{S_\beta}^{\alpha\gamma}\partial_{\alpha}T
f_{_{TT}}+[\frac{1}{h}\partial_{\alpha}(h{h_a}^{\beta}{S_\beta}^{\alpha\gamma})
+{h_a}^{\beta}{T^\lambda}_{\alpha\beta}{S_\lambda}^{\gamma\alpha}]f_{_T}
+\frac{1}{4}{h_a}^{\gamma}f=\frac{1}{2}\kappa^{2}{h_a}^{\beta}\Theta^{\gamma}_{\beta},
\end{equation}
where subscript $T$ and $TT$ represent first and second order
derivatives of $f$ with respect to $T$. In terms of covariant
derivative instead of partial derivatives, the $f(T)$ field
equations can be reconstructed. In covariant formalism, we obtain an
important condition, $R=-T-2\nabla^{\beta}{T^\gamma}_{\beta\gamma}$.
This implies that the covariant derivative of torsion tensor depicts
the only difference between Ricci and torsion scalars.

By implying the strategy of covariant derivative (widely discussed
in (Sharif \& Rani 2013, 2014a,b; Jawad \& Rani 2015), we get the
field equations in $f(T)$ gravity as
\begin{equation}\label{118a}
f_{T}G_{\alpha\gamma}+\frac{1}{2}g_{\alpha\gamma}(f-Tf_T)+
\mathcal{D}_{\alpha\gamma}f_{TT}=\kappa^2\Theta_{\alpha\gamma},
\end{equation}
where
$\mathcal{D}_{\alpha\gamma}={S_{\gamma\alpha}}^{\beta}\nabla_{\beta}T$.
The trace of the above equation is
\begin{equation}\label{119a}
\mathcal{D}f_{TT}-(R+2T)f_T+2f=\kappa^2\Theta,
\end{equation}
with $\mathcal{D}={\mathcal{D}^\gamma}_{\gamma}$ and
$\Theta={\Theta^{\gamma}}_{\gamma}$. Equation (\ref{118a}) can be
rewritten as
\begin{equation}\label{120a}
G_{\alpha\gamma}=\frac{\kappa^2}{f_{_T}}(\Theta_{\alpha\gamma}^{m}+\Theta_{\alpha\gamma}^{T}).
\end{equation}
The $\Theta_{\alpha\gamma}^{m}$ constitutes the matter contribution
while torsion contribution is presented by
\begin{equation}\label{122a}
\Theta_{\alpha\gamma}^{T}=\frac{1}{\kappa^2}[-\mathcal{D}_{\alpha\gamma}f_{_{TT}}-
\frac{1}{4}g_{\alpha\gamma}(\Theta-\mathcal{D}f_{_{TT}}+Rf_{_T})].
\end{equation}
It can be observed that (\ref{118a}) has an equivalent structure
such as $f(R)$ gravity and reduces to GR for $f=T$.

We consider the collapsing star as cylindrical symmetric surface
which is characterized by a hypersurface $\Sigma$. This timelike
3$D$ hypersurface isolates manifold into 4$D$ interior and exterior
portions. The interior region is taken as cylindrical symmetric
collapsing star having line element as follows
\begin{equation}\label{1}
ds^{2(-)}=X^2(t,r)dt^{2}-Y^2(t,r)dr^{2}-Z^2(t,r)d\phi^2-dz^2,
\end{equation}
where the coordinates $t,~r,~\phi$ and $z$ are constrained such as
\begin{equation*}
-\infty\leq t\leq \infty,\quad 0 \leq r< \infty, \quad 0\leq
\phi\leq 2\pi,\quad -\infty< z<\infty.
\end{equation*}
in order to preserve cylindrical symmetry. The line element for
exterior portion in terms of retarded time coordinate $\tau$ and
gravitational mass $M$ is given by (Chao-Guang 1995)
\begin{equation}\label{2}
ds^{2(+)}=\left(-\frac{2M}{r}\right)d\tau^2+2drd\tau-r^2(d\phi^2+\gamma^2
dz^2),
\end{equation}
where $\gamma$ is an arbitrary constant and $M=M(\tau)$ represents
total mass inside the cylindrical surface. Also, we take the
interior region is distributed by locally anisotropic matter for
which energy-momentum description is
\begin{equation}\label{3}
\Theta^{m}_{\alpha\beta}=(\rho+P_{r})U_{\alpha}U_{\beta}-P_{r}g_{\alpha\beta}
+(P_z-P_{r})V_{\alpha}V_{\beta}+(P_\phi-P_{r})L_{\alpha}L_{\beta},
\end{equation}
where $\rho=\rho(t,r)$ denotes the energy density,
$P_r=P_r(t,r),~P_{z}=P_{z}(t,r),~P_{\phi}=P_{\phi}(t,r)$ represent
the corresponding principal pressure components. The four velocity,
$U_\alpha=X\delta^0_\alpha$, the four vectors,
$L_\alpha=Z\delta^2_\alpha$ and $V_\alpha=\delta^3_\alpha$ satisfy
the relations
\begin{equation*}
U_\alpha U^\alpha=1,~L_\alpha L^\alpha=1=V_\alpha V^\alpha,~U^\alpha
L_\alpha=L^\alpha V_\alpha=V^\alpha U_\alpha=0.
\end{equation*}

\section{Basic Equations}

The choice of tetrad field in $f(T)$ gravity is the key role to
setup the framework of $f(T)$ gravity. The bad choice of tetrad are
those which constrain the torsion scalar to be constant or vanishes
the modification of theory. These tetrad consist in form of diagonal
presentation except cartesian symmetry. For spherical and
cylindrical symmetry, the good tetrad are non-diagonal tetrad having
no restriction for torsion scalar and keeps the modification of the
theory. We consider the following tetrad in non-diagonal form for
the interior spacetime as
\begin{center}${h^a}_{\alpha}=\left(\begin{array}{cccc}
X & 0 & 0 & 0 \\
0 & Y\cos\phi & -Z\sin\phi & 0 \\
0 & Y\sin\phi & Z\cos\phi & 0 \\
0 & 0 & 0 & 1 \\
\end{array}\right)$
\end{center}
with its inverse
\begin{center}${h_a}^{\alpha}=\left(\begin{array}{cccc}
\frac{1}{X} & 0 & 0 & 0 \\
0 & \frac{\cos\phi}{Y} & \frac{\sin\phi}{Y} & 0 \\
0 & -\frac{\sin\phi}{Z} & \frac{\cos\phi}{Z} & 0 \\
0 & 0 & 0 & 1 \\
\end{array}\right).$
\end{center}
The torsion scalar takes the form
\begin{eqnarray}\nonumber
T&=&-2\left[\frac{X'}{XY^2}\left(\frac{Y}{Z}-\frac{Z'}{Z}\right)
+\frac{\dot{Y}\dot{Z}}{X^2YZ}\right],
\end{eqnarray}
where dot and prime denotes derivative with respect to time and
radial coordinate. The corresponding field equations are
\begin{eqnarray}\nonumber
&&\frac{X^2}{Y^2}\left(\frac{Y'Z'}{YZ}-\frac{Z''}{Z}\right)+\frac{\dot{Y}\dot{Z}}{YZ}
=\frac{X^{2}\kappa^2}{f_{_T}}\left[{\rho}+
\frac{1}{\kappa^2}\left\{\frac{Tf_{_T}-f}{2}+\frac{1}{2Y^2}\left(\frac{Y}{Z}\right.\right.
\right.\\\label{4}&&-\left.\left.\left.
\frac{Z'}{Z}\right)f_{_T}'\right\}\right],
\\\label{5}
&&\frac{\dot{Z}'}{Z}-\frac{X'\dot{Z}}{XZ}-\frac{\dot{Y}Z'}{YZ}
=\frac{\dot{Z}}{2Z}\frac{f_{_T}'}{f_{_T}},\\\label{6}
&&\frac{\dot{Z}'}{Z}-\frac{X'\dot{Z}}{XZ}-\frac{\dot{Y}Z'}{YZ}
=\frac{1}{2}\left(\frac{Z'}{Z}-\frac{Y}{Z}\right)\frac{\dot{T}}{T'}\frac{f_{_T}'}{f_{_T}},\\
\nonumber
&&\frac{Y^2}{X^2}\left(\frac{\dot{X}\dot{Z}}{XZ}-\frac{\ddot{Z}}{Z}\right)+
\frac{X'Z'}{XZ}=\frac{Y^{2}\kappa^2}{f_{_T}}\left[P_r+
\frac{1}{\kappa^2}\left\{\frac{f-Tf_{_T}}{2}\right.\right.\\\label{9}&&-
\left.\left.-\frac{\dot{Z}\dot{T}}{2X^2ZT'}
f_{_T}'\right\}\right],\\\nonumber
&&\frac{Z^2}{XY}\left[\frac{X''}{Y}-\frac{\ddot{Y}}{X}+\frac{\dot{X}\dot{Y}}{X^2}
-\frac{X'Y'}{Y^2}\right]=\frac{Z^{2}\kappa^2}{f_{_T}}\left[P_\phi+
\frac{1}{\kappa^2}\left\{\frac{f-Tf_{_T}}{2}\right.\right.\\\label{7}&&-
\left.\left.\left(\frac{\dot{Y}\dot{T}}{2X^2YT'}
-\frac{X'}{2XY^2}\right)f_{_T}'\right\}\right],\\\nonumber
&&\frac{1}{Y^2} \left[\frac{X''}{X}-\frac{X'Y'}{XY}+\frac{X'Z'}{XZ}
+\frac{Z''}{Z}-\frac{Y'Z'}{YZ}\right]+\frac{1}{X^2}
\left[-\frac{\ddot{Y}}{Y}+\frac{\dot{X}\dot{Y}}{XY}-\frac{\ddot{Z}}{Z}
\right.\\\nonumber&&
\left.+\frac{\dot{X}\dot{Z}}{XZ}-\frac{\dot{Y}\dot{Z}}{YZ}\right]
=\frac{\kappa^2}{f_{_T}}\left[P_z+
\frac{1}{\kappa^2}\left\{\frac{f-Tf_{_T}}{2}-\left(\frac{\dot{T}}{2X^2T'}
\left(\frac{\dot{Y}}{Y}+\frac{\dot{Z}}{Z}\right)\right.\right.\right.\\\label{7}&&
\left.\left.\left.-\frac{1}{2Y^2}
\left(\frac{X'}{X}-\frac{Y}{Z}+\frac{Z'}{Z}\right)\right)f_{_T}'\right\}\right].
\end{eqnarray}
We find a relationship using Eqs.(\ref{5}) and (\ref{6}) as follows
\begin{eqnarray}\label{xyc5}
\frac{\dot{Z}}{Z}=\frac{\dot{T}}{T'}\bigg(\frac{Z'}{Z}-\frac{Y}{Z}\bigg).
\end{eqnarray}

In order to match interior and exterior regions of cylindrical
symmetric collapsing star, we use junction conditions defined by
Darmois. For this purpose, we consider the C-energy, i.e., mass
function representing matter inside the cylinder is given by (Throne
1965)
\begin{equation}
m(t,r)= \frac{l}{8}\bigg(1-\frac{1}{l^2}\nabla^\alpha
\hat{r}\nabla_\alpha\hat{r}\bigg)=E(t,r),
\end{equation}
where $E$ is the gravitational energy per unit specific length $l$
of the cylinder. The areal radius $\hat{r}$ is defined as
$\hat{r}=\mu l$ where circumference radius has the relation
$\mu^2=\xi_{(1)i}\xi_{(1)}^{i}$ while $l^2=\xi_{(2)i}\xi_{(2)}^{i}$.
The terms $\xi_{(1)}=\frac{\partial}{\partial \phi}$ and
$\xi_{(2)}=\frac{\partial}{\partial z}$ are the Killing vectors
corresponding to cylindrical systems. For Eq.(\ref{1}), the C-energy
turns out
\begin{equation}\label{10}
m(t,r)=\frac{l}{8}\bigg[1+\bigg(\frac{\dot{Z}}{X}\bigg)^2-\bigg(\frac{Z'}{Y}\bigg)^2\bigg].
\end{equation}
The continuity of Darmois conditions (junction conditions)
establishes the following constraints
\begin{eqnarray}\label{31}
\frac{l}{8}\overset{\Sigma^{(e)}}{=}m(t,r)-M, \quad
l\overset{\Sigma^{(e)}}{=}4C, \quad
-p_r\overset{\Sigma^{(e)}}{=}\frac{\Theta^{^T}_{11}}{Y^2}
-\frac{\Theta^{^T}_{01}}{XY}
\end{eqnarray}
where $\Sigma^{(e)}$ indicates the outer region for measurements
with $r=r_{\Sigma^{(e)}}$=constant. The conservation of total energy
of a system is obtained from Bianchi identities through dynamical
equations. The dynamical equations in the framework of $f(T)$
gravity are
\begin{eqnarray}\label{18c5}
\left(\overset{m}{\Theta^{\alpha\beta}}+\overset{T}{\Theta^{\alpha\beta}}\right)_{;~\beta}
U_{\alpha}=0,\quad
\left(\overset{m}{\Theta^{\alpha\beta}}+\overset{T}{\Theta^{\alpha\beta}}\right)_{;~\beta}
\chi_{\alpha}=0.
\end{eqnarray}
Using these equations, the dynamical equations for the cylindrically
symmetric collapsing star become
\begin{eqnarray}\label{11}
\dot{\rho}+\frac{\dot{Y}}{Y}(\rho+P_r)+\frac{\dot{Z}}{Z}(\rho+P_\phi)+\frac{J_{_0}}{\kappa^2}
=0,\\\label{12}
{P_r}'+\frac{X'}{X}(\rho+P_r)+\frac{Z'}{Z}(P_r-P_\phi)+\frac{J_{_1}}{\kappa^2}
=0,
\end{eqnarray}
where
\begin{eqnarray}\nonumber
J_{_0}&=&X^2\bigg\{\frac{1}{X^2}\bigg(\frac{Tf_{_T}-f}{2}+\frac{1}{2Y^2}\bigg(\frac{Y}{Z}
-\frac{Z'}{Z}\bigg)f_{_T}'\bigg)\bigg\}_{,0}+X^2\bigg\{\frac{\dot{Z}}{2X^2Y^2Z}f_{_T}'
\bigg\}_{,1}
\\\nonumber&+&\frac{\dot{X}}{X}\bigg(Tf_{_T}-f\bigg)+\bigg\{\frac{1}{YZ}
\bigg(\frac{\dot{X}}{X}
-\frac{\dot{X}Z'}{XY}+\frac{3X'\dot{Z}}{2XY}+\frac{\dot{Y}}{2Y}+\frac{\dot{Z}}{2Z}-
\frac{\dot{Y}Z'}{2Y^2}\\\nonumber&+&\frac{Y'\dot{Z}}{2Y^2}\bigg)+\bigg(\frac{X'Z'}
{2XY^2Z}-\frac{X'Y}{2XY^2Z}\bigg)\frac{\dot{T}}{T'}\bigg\}f_{_T}',\\\nonumber
J_{_1}&=&Y^2\bigg\{\frac{1}{2X^2Y^2}\bigg(\frac{Z'}{Z}-\frac{Y}{Z}\bigg)
\frac{\dot{T}}{T'}f_{_T}'
\bigg\}_{,0}+Y^2\bigg\{\frac{1}{Y^2}\bigg(\frac{f-Tf_{_T}}{2}-\frac{\dot{Z}\dot{T}}{2X^2ZT'}
f_{_T}'\bigg)\bigg\}_{,1}\\\nonumber&+&\frac{Y'}{Y}\bigg(f-Tf_{_T}\bigg)
+\bigg\{\bigg(\frac{3\dot{Y}Z'}{2X^2YZ}
-\frac{\dot{Y}}{X^2Z}-\frac{Y'\dot{Z}}{X^2YZ}+\frac{X'}{2XYZ}-\frac{X'Z'}{2XY^2Z}\\\nonumber&+&
\frac{\dot{X}Z'}{2X^3Z}-\frac{Y\dot{X}}{2X^3Z}-\frac{Y\dot{Z}}{2X^2Z^2}
-\frac{X'Z'}{2X^2Z}\bigg)\frac{\dot{T}}{T'}-\frac{X'Z'}{XY^2Z}+
\frac{\dot{Y}\dot{Z}}{2X^2YZ}\bigg\}f_{_T}'.
\end{eqnarray}

In order to develop the collapse equation representing the dynamics
of cylindrically symmetric collapsing star in the framework of
$f(T)$ gravity, the choice of $f(T)$ model keeps central importance.
Here we assume model in form of power-law upto quadratic torsion
scalar term to discuss the instability ranges of cylindrical
symmetric collapsing star. This is given by (Sharif and Rani 2014,
2015; Jawad and Rani 2015)
\begin{equation}\label{13}
f(T)=T+\omega T^2,
\end{equation}
where $\omega$ is an arbitrary constant. This particular model is
analogues to a viable model from $f(R)$ gravity such as
$f(R)=R+\lambda R^2$ which reduces to GR by taking
$\lambda\rightarrow 0$. In $f(R)$ gravity, this model takes part to
present dynamics of collapsing star and gives instability ranges for
different regimes under many scenarios (Sharif and Kausar 2011,
Kausar and Noureen 2014). The power-law $f(T)$ model is very simple
form to be considered which provides direct comparison with GR by
choosing $\omega$ as zero. The finite time singularities are also
discussed for power-law model of the type $T^m$ which results the
vanishing of these singularities for $m>1$ (Bamba et al. 2012).
Also, this model leads to the accelerated expansion of the universe
in phantom phase, possibility of realistic wormhole solutions, solar
system tests and instability conditions for a collapsing star. In
order to discuss the dynamical instability ranges in the underlying
scenario, we impose the condition of static equilibrium (dependent
of radial coordinate only) on metric, matter as well as effective
parts of the system initially. After some time $t$, these parts also
become time dependent along with $r$ dependency. We represent this
strategy by linear perturbation strategy to construct the dynamical
equations in order to explore instability ranges for the
cylindrically symmetric collapsing star. These perturbations are
described as follows
\begin{eqnarray}\label{41cc5}
X(t,r)&=&x_0(r)+\varepsilon \Lambda(t)\hat{x}(r),\\\label{42cc5}
Y(t,r)&=&y_0(r)+\varepsilon \Lambda(t)\hat{y}(r),\\\label{43cc5}
Z(t,r)&=&z_0(r)+\varepsilon \Lambda(t)\hat{z}(r),\\\label{44cc5}
\rho(t,r)&=&\rho_0(r)+\varepsilon {\hat{\rho}}(t,r),\\\label{45cc5}
P_r(t,r)&=&p_{r0}(r)+\varepsilon {\hat{p}_r}(t,r), \\\label{46cc5}
P_{\phi}(t,r)&=&p_{\phi0}(r)+\varepsilon{\hat{p}_\phi}(t,r),\\\label{48cc5}
P_{z}(t,r)&=&p_{z0}(r)+\varepsilon{\hat{p}_z}(t,r),\\\label{48cc5}
m(t,r)&=&m_0(r)+\varepsilon \hat{m}(t,r), \\\label{49'cc5}
T(t,r)&=&T_0(r)+\varepsilon \Lambda(t)e(r).
\end{eqnarray}
That is, the quantities with zero subscript refer zero order
perturbation of corresponding functions while $0<\varepsilon\ll1$.
The $f(T)$ model under perturbation of torsion scalar becomes
\begin{eqnarray}\label{a+}
f(T)=T_0(1+\omega T_0)+\varepsilon \Lambda e(1+2\omega T_0),\quad
f_{_T}(T)=1+2\omega T_0+2\varepsilon \delta \Lambda e.
\end{eqnarray}

\section{Collapse Equation}

Here, we construct the collapse equation using $f(T)$ model along
with perturbation scheme for the underlying scenario. The static
configuration of torsion scalar and mass function with $Z_0=r$ are
given as follows
\begin{equation*}
T_0=-\frac{2x_0'(y_o-1)}{rx_0y_0},\quad
m_0=\frac{l}{8}\left(1-\frac{1}{y_0^2}\right),
\end{equation*}
while perturbed configuration is given by
\begin{eqnarray*}
e&=&-\frac{2}{rx_0y_0}\bigg[x_0'\left(\hat{y}-\hat{z}'+\frac{\hat{z}}{r}(1-y_0)\right)
+(y_0-1)\left(\hat{x}'-\frac{x_0'}{x_0y_0}(\hat{x}y_0+\hat{y}x_0)\right)\bigg],\\
\hat{m}&=&-\frac{\Lambda
l}{4y_0^2}\left(\hat{z}'-\frac{\hat{y}}{y_0}\right),
\end{eqnarray*}
respectively. The dynamical equations take part in order to
construct collapse equation for the dynamical instability ranges of
cylindrical collapsing star. For this purpose, the non-static
configuration of second dynamical equation (\ref{12}) takes the form
\begin{eqnarray}\nonumber
\hat{p}_{r}'+\frac{x_0'}{x_0}(\hat{\rho}+\hat{p}_r)+\bigg(\frac{\hat{x}'}{x_0}
-\frac{\hat{x}x_0'}{x_0}\bigg)({\rho}_0+p_{r0})\Lambda+\frac{1}{r}(\hat{p}_r-\hat{p}_\phi)
\\\label{20}+\frac{\hat{z}'}{r}(p_{r0}-p_{\phi0})\Lambda
+\frac{J_{1p}}{\kappa^2}=0,
\end{eqnarray}
where
\begin{eqnarray}\nonumber
J_{1p}&=&\frac{e\omega
T_0^{2'}}{rx_0^2}(1-y_0)\ddot{\Lambda}+2y_0\hat{y}\bigg(\frac{\omega
T_0^{2}}{2y_0^2}\bigg)_{,1}\Lambda+y_0^2\bigg\{\frac{1}{y_0^2}
\bigg(\frac{\hat{y}\omega T_0^{2}}{y_0}\\\nonumber&-&\omega e
T_0\bigg)\bigg\}_{,1}\Lambda-\frac{2\omega eT_0
y_0'}{y_0}\Lambda-\frac{1}{y_0}\bigg(\hat{y}'-\frac{y_0'\hat{y}}{y_0}\bigg)\omega
T_0^2\Lambda+\frac{e\omega
T_0^{2'}}{rx_0}\bigg(\frac{x_0'}{y_0}\\\nonumber&-&\frac{\hat{x}'}{y_0^2}
-\frac{\hat{x}'}{x_0}\bigg) -\frac{2\hat{x}'\omega
e'}{rx_0y_0^2}\Lambda-2\omega
T_0'\bigg(\frac{\hat{x}}{x_0}+\frac{\hat{y}}{y_0}+\frac{\hat{z}}{r}\bigg)\Lambda.
\end{eqnarray}
Equation (\ref{20}) is the general collapse equation which depicts
the instability of hydrostatic equilibrium of cylindrical
gravitating fluid in $f(T)$ gravity. To analyze the instability of
fluid using this equation, we need the expressions for
$\hat{\rho},~\hat{p}_r,~\hat{p}_\phi$ and $\Lambda$ through basic
equations of the underlying system.

Applying perturbation strategy on first dynamical equation, the
perturbed part is given as follows
\begin{eqnarray}\label{j4c5}
\dot{\hat{\rho}}+\left[\frac{\hat{y}}{y_0}(\rho_0+p_{r0})+\frac{\hat{z}}{r}(\rho_0+p_{\phi0})
+\frac{J_{0p}}{\kappa^2}\right]\dot{\Lambda}=0,
\end{eqnarray}
where
\begin{eqnarray}\nonumber
J_{0p}&=&e\omega T_0+\frac{\omega e'(y_0-1)}{ry_0^2}+\frac{\omega
T_0'}{ry_0^2}\bigg(\hat{y}-\hat{z}'+\frac{\hat{z}}{r}(1-y_0)-\frac{2\hat{y}(y_0-1)}{y_0}\bigg)
\\\nonumber&-&\frac{\hat{x}}{x_0}\bigg(\omega T_0^2+\frac{2\omega T_0'(y_0-1)}{ry_0^2}\bigg)
+x_0^2\bigg(\frac{z\omega
T_0'}{rx_0^2y_0^2}\bigg)+\frac{\hat{x}\omega
T_0^2}{x_0}+\frac{2\omega
T_0'}{ry_0}\\\nonumber&\times&\bigg(\frac{\hat{x}}{x_0}-\frac{\hat{x}}{x_0y_0}+
\frac{\hat{z}x_0'}{2x_0y_0}+\frac{\hat{y}}{2y_0}+\frac{\hat{z}}{2r}-
\frac{\hat{y}}{2y_0^2}+\frac{\hat{z}x_0'}{x_0y_0}+\frac{\hat{z}y_0'}{2y_0^2}
\\\nonumber&+&\frac{ex_0'T_0'(1-y_0)}{2x_0y_0}\bigg)
\end{eqnarray}
Integrating Eq.(\ref{j4c5}) with respect to time, we obtain the
non-static energy density which is given by
\begin{eqnarray}\label{62cc5}
\hat{\rho}&=&-\left[\frac{\hat{y}}{y_0}(\rho_0+p_{r0})+\frac{\hat{z}}{r}(\rho_0+p_{\phi0})
+\frac{J_{0p}}{\kappa^2}\right]\Lambda.
\end{eqnarray}

The Harrison-Wheeler equation of state establishes a relationship
between energy density and pressure such as (Harrison et al. 1965)
\begin{equation}\label{j7c5}
{\hat{p}_{i}}=\Gamma \frac{p_{i0}}{\rho_0+p_{i0}}\hat{\rho},
\end{equation}
where $\Gamma$ is called adiabatic index. We use this index in order
to examine instability ranges in the context of $f(T)$ gravity. The
adiabatic index $\Gamma$ finds the rigidity of the fluid and
evaluates the change of pressure to corresponding density.
Substituting the value of $\hat{\rho}$ from Eq.(\ref{62cc5}) in
(\ref{j7c5}) for $\hat{p}_r,$ and $\hat{p}_\phi$, it yields
\begin{eqnarray}\label{j8c5}
{\hat{p}_r}&=&-\Lambda
\left[\frac{\hat{y}}{y_0}p_{r0}+\frac{c}{r}\frac{\rho_0+p_{\phi0}}{\rho_0
+p_{r0}}p_{r0}+\frac{1}{\kappa^2}\frac{p_{r0}}{\rho_0+p_{r0}}J_{0p}\right]\Gamma,
\\\label{j88c5}{\hat{p}_\phi}&=&-\Lambda
\left[\frac{\hat{y}}{y_0}\frac{\rho_0+p_{r0}}{\rho_0
+p_{\phi0}}p_{\phi0}+\frac{c}{r}p_{\phi0}+\frac{1}{\kappa^2}
\frac{p_{\phi0}}{\rho_0+p_{\phi0}}J_{0p}\right]\Gamma.
\end{eqnarray}
To find out the value of $\Lambda(t)$, we perturbed field equation
(\ref{9}) and its non-static part is given by
\begin{eqnarray}\nonumber
-\frac{\hat{z}}{rx_0^2}\ddot{\Lambda}+\frac{1}{rx_0y_0^2}\bigg(-\frac{2\hat{y}x_0'}{y_0}
-\frac{\hat{x}x_0'}{x_0}-\frac{\hat{z}x_0'}{r}+\hat{x}'+\hat{z}'x_0'\bigg)\Lambda=\\\label{30}
-\frac{2\omega e\kappa^2\Lambda}{(1+2\omega
T_0)^2}p_{r0}+\frac{\kappa^2 \hat{p}_r}{1+2\omega T_0}-\frac{e
\omega T_0 \Lambda}{1+2\omega T_0}-\frac{e \omega^2 T_0^2
\Lambda}{(1+2\omega T_0)^2}.
\end{eqnarray}

We use the last junction condition given in Eq.(\ref{31}) with
$r=r_{_\Sigma^{(e)}}$=constant which under perturbation strategy
yields
\begin{eqnarray}\label{32}
p_{r0}\overset{\Sigma^{(e)}}{=}\frac{\omega T_0^2}{2\kappa^2},\quad
\hat{p}_r\overset{\Sigma^{(e)}}{=}\frac{\omega e
T_0}{\kappa^2}\Lambda.
\end{eqnarray}
Using this equation along with $r=r_{_\Sigma^{(e)}}$=constant in
Eq.(\ref{30}), we obtain
\begin{equation}\nonumber
\ddot{\Lambda}-\sigma_{\Sigma^{(e)}}\Lambda\overset{\Sigma^{(e)}}{=}0,\quad
\textmd{where}\quad
\sigma_{\Sigma^{(e)}}=\frac{2r_{\Sigma^{(e)}}\omega^2 e
T_0^2x_0^2}{\hat{z}(1+2\omega T_0)^2}.
\end{equation}
Its solution is
\begin{equation*}
\Lambda(t)=c_1 e^{\sqrt{\sigma_{\Sigma^{(e)}}}~t}+c_2
e^{-\sqrt{\sigma_{\Sigma^{(e)}}}~t},
\end{equation*}
representing stability and instability phases of cylindrical
gravitating fluid through static and non-static parts and $c_1,
~c_2$ are constants. We assume the hydrostatic equilibrium for which
$\Lambda$ is zero at large past time, $t=-\infty$. After this
scenario with the evolution of time, the system recruits into
present phase, reduce its areal radius and commences to collapse. In
regards to discuss instability analysis of cylindrical collapsing
star, we take only static solution ($\Lambda(-\infty)=0$) which
gives $c_2=0$. Choosing $c_1=-1$, we get
\begin{eqnarray}\label{33}
\Lambda(t)=- e^{\sqrt{\sigma_{\Sigma^{(e)}}}~t},\quad
\textmd{where}\quad \sigma_{\Sigma^{(e)}}>0.
\end{eqnarray}
Inserting all the corresponding values in general collapse equation
(\ref{20}), we obtain
\begin{eqnarray}\nonumber
&&\Gamma\bigg[\bigg\{\frac{\hat{y}}{y_0}p_{r0}+\frac{c}{r}\frac{\rho_0+p_{\phi0}}{\rho_0
+p_{r0}}p_{r0}+\frac{1}{\kappa^2}\frac{p_{r0}}{\rho_0+p_{r0}}J_{0p}\bigg\}_{,1}
+\frac{x_0'}{x_0}\bigg\{\frac{\hat{y}}{y_0}p_{r0}\\\nonumber&&+\frac{c}{r}
\frac{\rho_0+p_{\phi0}}{\rho_0
+p_{r0}}p_{r0}+\frac{1}{\kappa^2}\frac{p_{r0}}{\rho_0+p_{r0}}J_{0p}\bigg\}+\frac{1}{r}
\bigg\{\frac{\hat{y}}{y_0}p_{r0}+\frac{c}{r}\frac{\rho_0+p_{\phi0}}{\rho_0
+p_{r0}}p_{r0}\\\nonumber&&+\frac{1}{\kappa^2}\frac{p_{r0}}{\rho_0+p_{r0}}J_{0p}-
\bigg(\frac{\hat{y}}{y_0}\frac{\rho_0+p_{r0}}{\rho_0
+p_{\phi0}}p_{\phi0}+\frac{c}{r}p_{\phi0}+\frac{1}{\kappa^2}
\frac{p_{\phi0}}{\rho_0+p_{\phi0}}J_{0p}\bigg)\bigg\}\bigg]\\\nonumber&&+
\frac{x_0'}{x_0}\bigg\{\frac{\hat{y}}{y_0}(\rho_0+p_{r0})+\frac{\hat{z}}{r}(\rho_0+p_{\phi0})
+\frac{J_{0p}}{\lambda^2}\bigg\}-\bigg(\frac{\hat{x}'}{x_0}
-\frac{\hat{x}x_0'}{x_0}\bigg)({\rho}_0+p_{r0})\\\label{35}&&-
\frac{\hat{z}'}{r}(p_{r0}-p_{\phi0})
+\frac{J_{1p}}{\kappa^2}\frac{1}{e^{\sqrt{\sigma_{\Sigma^{(e)}}}~t}}=0.
\end{eqnarray}
This equation represents the collapse equation for cylindrically
gravitating object in the framework of $f(T)$ gravity.

\section{Dynamical Instability Analysis}

In this section, we analyze the instability ranges of cylindrically
symmetric self-gravitating fluid in $f(T)$ gravity using Newtonian
and post-Newtonian conditions in the collapse equation for
corresponding regimes with the help of adiabatic index.

\subsection{Newtonian Regime}

For Newtonian regime, we have the constraints as follows
\begin{equation*}
x_0=1=y_0,\quad \rho_0\geq p_{r0},\quad \rho_0\geq p_{\phi0}.
\end{equation*}
These constraints implies that
$x_0'=0=y_0',~\frac{p_{r0}}{\rho_0+p_{r0}}\rightarrow0$ and
$\frac{p_{\phi0}}{\rho_0+p_{\phi0}}\rightarrow0$. Using these
expressions in Eq.(\ref{35}) and consequently collapse equation
reduces to
\begin{eqnarray}\nonumber
\Gamma\bigg[\bigg(\hat{y}p_{r0}+\frac{z}{r}p_{r0}\bigg)_{,1}+\frac{1}{r}\bigg(\hat{y}
+\frac{\hat{z}}{r}\bigg)
(p_{r0}-p_{\phi0})\bigg]-\hat{x}'\rho_0-\frac{\hat{z}'}{r}(p_{r0}-p_{\phi0})
\\\label{36}+\frac{J_{1p}^{^N}}{\kappa^2}\frac{1}{e^{\sqrt{\sigma_{\Sigma^{(e)}}}~t}}=0
\end{eqnarray}
which represents the cylindrically symmetric self-gravitating fluid
in hydrostatic equilibrium. Here $J_{1p}^{^N}$ expresses Newtonian
approximation terms in $J_{1p}$, i.e., those terms remained after
applying above constraints. The system turns to collapses or
unstable if
\begin{eqnarray}\label{37}
\Gamma<\frac{\hat{x}'\rho_0+\frac{\hat{z}'}{r}(p_{r0}-p_{\phi0})-
\frac{J_{1p}^{^N}}{\kappa^2}\frac{1}{e^{\sqrt{\sigma_{\Sigma^{(e)}}}~t}}}
{\bigg(\hat{y}p_{r0}+\frac{z}{r}p_{r0}\bigg)_{,1}+\frac{1}{r}\bigg(\hat{y}
+\frac{\hat{z}}{r}\bigg) (p_{r0}-p_{\phi0})}=\frac{A_N}{B_N}.
\end{eqnarray}

It is noted that we have assigned numerator and denominator as $A_n$
and $B_n$ for the sake for simplicity. We assume adiabatic index to
be positive throughout the scenario to maintain difference between
gradient of principal pressure components and gravitational forces.
Under this condition, the left hand side of above inequality
depending on dynamical properties such as density, pressure
components and torsion terms remains positive. The system remains
unstable as long as this inequality holds.  Thus it leads to
following possibilities
\begin{itemize}
  \item I:~~~$\hat{x}'\rho_0+\frac{\hat{z}'}{r}(p_{r0}-p_{\phi0})-
\frac{J_{1p}^{^N}}{\kappa^2}\frac{1}{e^{\sqrt{\sigma_{\Sigma^{(e)}}}~t}}=
\bigg(\hat{y}p_{r0}+\frac{z}{r}p_{r0}\bigg)_{,1}+\frac{1}{r}\bigg(\hat{y}
+\frac{\hat{z}}{r}\bigg) (p_{r0}-p_{\phi0})$ or equivalently
$A_N=B_N$
  \item II:~~~$\hat{x}'\rho_0+\frac{\hat{z}'}{r}(p_{r0}-p_{\phi0})-
\frac{J_{1p}^{^N}}{\kappa^2}\frac{1}{e^{\sqrt{\sigma_{\Sigma^{(e)}}}~t}}<
\bigg(\hat{y}p_{r0}+\frac{z}{r}p_{r0}\bigg)_{,1}+\frac{1}{r}\bigg(\hat{y}
+\frac{\hat{z}}{r}\bigg) (p_{r0}-p_{\phi0})$ or equivalently
$A_N<B_N$
  \item III:~~~$\hat{x}'\rho_0+\frac{\hat{z}'}{r}(p_{r0}-p_{\phi0})-
\frac{J_{1p}^{^N}}{\kappa^2}\frac{1}{e^{\sqrt{\sigma_{\Sigma^{(e)}}}~t}}>
\bigg(\hat{y}p_{r0}+\frac{z}{r}p_{r0}\bigg)_{,1}+\frac{1}{r}\bigg(\hat{y}
+\frac{\hat{z}}{r}\bigg) (p_{r0}-p_{\phi0})$ or equivalently
$A_N>B_N$
\end{itemize}
In the case I, we obtain $\Gamma<\frac{A_N}{B_N}=1$ which implies
that $0<\Gamma<1$, the instability range for cylindrical system in
Newtonian range. For the case II, the condition $A_N<B_N$ yields
$\Gamma<\frac{A_N}{B_N}<1$ which leads to the instability range as
$0<\Gamma<\frac{A_N}{B_N}$ where $\frac{A_N}{B_N}<1$. If $A_N>B_N$
holds, we obtain the instability range by the inequality as
$0<\Gamma<\frac{A_N}{B_N}$ where $\frac{A_N}{B_N}>1$. It is noted
that, we may recover the GR in Newtonian limit for instability range
as $\Gamma<\frac{4}{3}$ in this case. This limit also includes the
first two instability ranges.\\\\ {\textbf{Isotropic Pressure}}\\\\
Here we discuss the instability ranges for the case when all
pressure components become equal, i.e., isotropic pressure fluid
$(p_r=p_\phi=p_z=p)$. For isotropic cylindrical fluid system,
Eq.(\ref{37}) reduces to
\begin{eqnarray}\label{38}
\Gamma<\frac{\hat{x}'\rho_0-
\frac{J_{1p}^{^N}}{\kappa^2}\frac{1}{e^{\sqrt{\sigma_{\Sigma^{(e)}}}~t}}}
{(\hat{y}p+\frac{\hat{z}}{r}p)_{,1}}.
\end{eqnarray}
This inequality indicates the unstable behavior if it holds and
expresses the correspondingly instability ranges.
\\\\ {\textbf{Asymptotic Behavior}}\\\\
Substituting $\omega=0$ in Eq.(\ref{37}) implies $J_{1p}^{^N}=0$, it
yields the inequality for anisotropic pressure fluid as
\begin{eqnarray}\label{39}
\Gamma<\frac{\hat{x}'\rho_0+\frac{\hat{z}'}{r}(p_{r0}-p_{\phi0})}
{\bigg(\hat{y}p_{r0}+\frac{z}{r}p_{r0}\bigg)_{,1}+\frac{1}{r}\bigg(\hat{y}
+\frac{\hat{z}}{r}\bigg) (p_{r0}-p_{\phi0})}.
\end{eqnarray}
The adiabatic index indicates the results for GR as long as above
inequality satisfied. For cylindrically symmetric isotropic fluid,
we obtain
\begin{eqnarray}\label{40}
\Gamma<\frac{\hat{x}'\rho_0} {(\hat{y}p+\frac{\hat{z}}{r}p)_{,1}}.
\end{eqnarray}
In these cases, the instability ranges are retrieved accordingly as
obtained for Eq.(\ref{37}).

\subsection{Post-Newtonian Regime}

Here we study the dynamical instability of cylindrical
self-gravitating object with post-Newtonian limits. These limits are
$x_0=1-\frac{m_0}{r},~y_0=1+\frac{m_0}{r}$ upto order
$O(\frac{m_0}{r})$. Consequently the fluid for hydrostatic
equilibrium by inserting $x_0$ and $y_0$ in collapse equation takes
the form
\begin{eqnarray}\nonumber
&&\Gamma\bigg[\bigg\{\hat{y}(1-\frac{m_0}{r})p_{r0}+\frac{c}{r}\frac{\rho_0+p_{\phi0}}{\rho_0
+p_{r0}}p_{r0}+\frac{1}{\kappa^2}\frac{p_{r0}}{\rho_0+p_{r0}}J_{0p}^{_{pN}}\bigg\}_{,1}
+\frac{m_0}{r^2}\bigg\{\hat{y}(1-\frac{m_0}{r})p_{r0}\\\nonumber&&+\frac{c}{r}
\frac{\rho_0+p_{\phi0}}{\rho_0
+p_{r0}}p_{r0}+\frac{1}{\kappa^2}\frac{p_{r0}}{\rho_0+p_{r0}}J_{0p}^{_{pN}}\bigg\}+\frac{1}{r}
\bigg\{\hat{y}(1-\frac{m_0}{r})p_{r0}+\frac{c}{r}\frac{\rho_0+p_{\phi0}}{\rho_0
+p_{r0}}p_{r0}\\\nonumber&&+\frac{1}{\kappa^2}\frac{p_{r0}}{\rho_0+p_{r0}}J_{0p}^{_{pN}}-
\bigg(\hat{y}(1-\frac{m_0}{r})\frac{\rho_0+p_{r0}}{\rho_0
+p_{\phi0}}p_{\phi0}+\frac{c}{r}p_{\phi0}+\frac{1}{\kappa^2}
\frac{p_{\phi0}}{\rho_0+p_{\phi0}}J_{0p}^{_{pN}}\bigg)\bigg\}\bigg]\\\nonumber&&+
\frac{m_0}{r^2}\bigg\{\hat{y}(\rho_0+p_{r0})+\frac{\hat{z}}{r}(\rho_0+p_{\phi0})
+\frac{J_{0p}^{_{pN}}}{\lambda^2}\bigg\}-\bigg(\hat{x}'(1+\frac{m_0}{r})
-\frac{\hat{x}m_0}{r^2}\bigg)({\rho}_0+p_{r0})\\\label{45}&&-
\frac{\hat{z}'}{r}(p_{r0}-p_{\phi0})
+\frac{J_{1p}^{_{pN}}}{\kappa^2}\frac{1}{e^{\sqrt{\sigma_{\Sigma^{(e)}}}~t}}=0.
\end{eqnarray}
The terms with superscript $pN$ point out the terms with
post-Newtonian approximations in corresponding expressions. For
dynamical instability, we find
\begin{eqnarray}\label{50}
\Gamma<\frac{A_{pN}}{B_{pN}},
\end{eqnarray}
where
\begin{eqnarray*}
A_{pN}&=&\bigg(\hat{x}'\bigg(1+\frac{m_0}{r}\bigg)
-\frac{\hat{x}m_0}{r^2}\bigg)({\rho}_0+p_{r0})-
\frac{m_0}{r^2}\bigg\{\hat{y}(\rho_0+p_{r0})-\frac{J_{0p}^{_{pN}}}{\kappa^2}
\\\nonumber&+&\frac{\hat{z}}{r}(\rho_0+p_{\phi0})
\bigg\}+ \frac{\hat{z}'}{r}(p_{r0}-p_{\phi0})
-\frac{J_{1p}^{_{pN}}}{\kappa^2}\frac{1}{e^{\sqrt{\sigma_{\Sigma^{(e)}}}~t}},\\\nonumber
B_{pN}&=&\bigg\{\hat{y}\bigg(1-\frac{m_0}{r}\bigg)p_{r0}+\frac{c}{r}
\frac{\rho_0+p_{\phi0}}{\rho_0
+p_{r0}}p_{r0}+\frac{1}{\kappa^2}\frac{p_{r0}}{\rho_0+p_{r0}}J_{0p}^{_{pN}}\bigg\}_{,1}
+\frac{m_0}{r^2}\\\nonumber&\times&\bigg\{\hat{y}p_{r0}+\frac{c}{r}
\frac{\rho_0+p_{\phi0}}{\rho_0
+p_{r0}}p_{r0}+\frac{1}{\kappa^2}\frac{p_{r0}}{\rho_0+p_{r0}}J_{0p}^{_{pN}}\bigg\}+\frac{1}{r}
\bigg\{\hat{y}\bigg(1-\frac{m_0}{r}\bigg)\\\nonumber&\times&p_{r0}+\frac{c}{r}
\frac{\rho_0+p_{\phi0}}{\rho_0
+p_{r0}}p_{r0}+\frac{1}{\kappa^2}\frac{p_{r0}}{\rho_0+p_{r0}}J_{0p}^{_{pN}}-
\bigg(\hat{y}\bigg(1-\frac{m_0}{r}\bigg)\\\nonumber&\times&\frac{\rho_0+p_{r0}}{\rho_0
+p_{\phi0}}p_{\phi0}+\frac{c}{r}p_{\phi0}+\frac{1}{\kappa^2}
\frac{p_{\phi0}}{\rho_0+p_{\phi0}}J_{0p}^{_{pN}}\bigg)\bigg\}.
\end{eqnarray*}
Similar to the case of Newtonian regime, we can develop three
possibilities on $A_{pN}$ and $B_{pN}$, i.e.,
$A_{pN}=B_{pN},~A_{pN}<B_{pN},~A_{pN}>B_{pN}$. These possibilities
yield the instability ranges as $0<\Gamma<\frac{A_{pN}}{B_{pN}}$
which contains $0<\Gamma<1$.
\\\\ {\textbf{Isotropic Pressure}}\\\\
Taking isotropy of pressure in Eq.(\ref{45}), we get the unstable
behavior as
\begin{eqnarray}\label{501}
\Gamma<\frac{A_{ipN}}{B_{ipN}},
\end{eqnarray}
where
\begin{eqnarray*}
A_{ipN}&=&\bigg(\hat{x}'\bigg(1+\frac{m_0}{r}\bigg)
-\frac{\hat{x}m_0}{r^2}\bigg)({\rho}_0+p)-
\frac{m_0}{r^2}\bigg\{\hat{y}(\rho_0+p)-\frac{J_{0p}^{_{pN}}}{\kappa^2}
\\\nonumber&+&\frac{\hat{z}}{r}(\rho_0+p)
\bigg\}
-\frac{J_{1p}^{_{pN}}}{\kappa^2}\frac{1}{e^{\sqrt{\sigma_{\Sigma^{(e)}}}~t}},\\\nonumber
B_{ipN}&=&\bigg\{\hat{y}\bigg(1-\frac{m_0}{r}\bigg)p_{r0}+\frac{c}{r}
p+\frac{1}{\kappa^2}\frac{p}{\rho_0+p}J_{0p}^{_{pN}}\bigg\}_{,1}
+\frac{m_0}{r^2}\\\nonumber&\times&\bigg\{\hat{y}p+\frac{c}{r}
p+\frac{1}{\kappa^2}\frac{p}{\rho_0+p}J_{0p}^{_{pN}}\bigg\}+\frac{1}{r}
\bigg\{\hat{y}\bigg(1-\frac{m_0}{r}\bigg)p+\frac{c}{r}
p+\frac{1}{\kappa^2}\\\nonumber&\times&\frac{p}{\rho_0+p}J_{0p}^{_{pN}}-
\bigg(\hat{y}\bigg(1-\frac{m_0}{r}\bigg)p+\frac{c}{r}p+\frac{1}{\kappa^2}
\frac{p}{\rho_0+p}J_{0p}^{_{pN}}\bigg)\bigg\}.
\end{eqnarray*}
The stability ranges for the cylindrically symmetric isotropic fluid
have the criteria as for inequality (\ref{50}).
\\\\ {\textbf{Asymptotic Behavior}}\\\\
In order to recover the results in GR, we insert $\omega=0$ which
leads to $J_{0p}^{_{pN}}=0=J_{1p}^{_{pN}}$ for isotropic as well as
anisotropic fluids.

\section{Conclusion}

The dynamical instability of a self-gravitating general relativistic
object undergoing gravitational collapse process has become widely
considered phenomena in GR as well as in modified theories of
gravity. This process stays significantly at the center of structure
formation and holds the evolutionary development of these objects.
We have taken cylindrically symmetric line element as interior
spacetime while exterior spacetime in retarded time coordinate in
the framework of $f(T)$ gravity. The locally anisotropic matter
distribution is considered for which we have obtained an important
result at boundary of matching both interior and exterior regions.
We have developed $f(T)$ field equations along with some basic
equations such that dynamical equations through Bianchi identities.
In order to get insights in more realistic way, we have assumed a
specific power-law $f(T)$ model with linear and quadratic torsion
scalar terms. This model is used to discuss many cosmological
scenarios as well as general symmetric solutions such as stability
of spherical collapsing star and wormhole solutions.

Keeping system in hydrostatic equilibrium initially and then
perturbed with the evolution of time by linear perturbation
strategy. This strategy is applied on all matter, metric, mass and
torsion components. In order to construct collapse equation, the
dynamical equations are used in an appropriate way along with
adiabatic index $\Gamma$. With the help of second law of
thermodynamics, this index gives the ratio of specific heat using
energy density and pressure components. We have obtained the
solution for non-static perturbed quantity using matching conditions
which satisfied the initial state of equilibrium. We have applied
the constraints of Newtonian and post-Newtonian regimes in collapse
equation to find the instability ranges for the cylindrically
symmetric collapsing object in the framework of $f(T)$ gravity.

We have found instability ranges for both regimes represented by
adiabatic index. We have also found these ranges taking into account
isotropic pressure as well as asymptotic behavior, i.e., GR
solutions. We have found the instability range as
$0<\Gamma<\frac{A}{B}$ for
$\frac{A}{B}=1,~\frac{A}{B}<1,~\frac{A}{B}>1$ where $A$ and $B$
represent the corresponding expressions in each case. These
expressions depend on matter, metric and torsion terms. This range
also admits the GR condition of unstable behavior through adiabatic
index which is $\Gamma<\frac{4}{3}$. It is noted that for other
forms of $f(T)$ models instead of power-law form, the quantitative
consequences are changed while qualitative consequences remain same.
For instance, we consider the exponential model,
$f(T)=T-\alpha_1T(1-e^{\frac{pT_c}{T}})$ for which we may obtain
static and perturbed parts using perturbation scheme adopting some
more steps. The collapse equation as well as instability ranges
through adiabatic index will depend on exponential terms throughout.
However, the qualitative consequences remain unchanged due to the
fact that all the instability ranges via adiabatic index depend on
matter, metric and torsion dependent terms.

The dynamical instability in $f(R)$ gravity has been discussed
taking CDTT model for a cylindrically symmetric collapsing star
(Kausar 2013). It has been found that adiabatic index depends on
immense perturbed terms of this model along with some positivity
constraints for the dynamical unstable behavior. In Brans-Dicke
theory of gravity (Sharif \& Manzoor 2015), the instability ranges
of a collapsing stellar object having cylindrically symmetry have
been investigated. It yields the instability ranges through collapse
equation depend on the dynamical variables of collapsing fluid. The
ranges for unstable behavior are obtained as  $0 < \Gamma < 1$ while
for a special case, it gives $\Gamma> 1$. In $f(T)$ gravity, we have
found dynamical instability ranges for spherically symmetric
collapsing star with and without expansion as well as with
shear-free conditions taking anisotropic fluid (Sharif \& Rani 2014,
2015; Jawad \& Rani 2015). However, in the present paper, we have
analyzed dynamical unstable behavior taking cylindrical symmetric
object which gives less complexity in expressions $A$ and $B$. Also,
we have reduced the results in the limit of isotropic fluid
distribution and to GR limit.

Chandrasekhar (1964) was the first who explored dynamical
instability ranges of a spherically symmetric isotropic fluid in GR.
He established these ranges through adiabatic index which depends on
its numerical value. That is, the weight of the outer layer
increases rapidly as compared to the pressure in a star for $\Gamma
<\frac{4}{3}$ which leads to the unstable behavior of the star.
Whereas, for $\Gamma >\frac{4}{3}$, the pressure overcomes the
weight of out layers and yields the stability of the star. In $f(T)$
gravity, Sharif \& Rani (2014, 2015) analyzed the dynamics of
self-gravitating object with spherical symmetry via expansion and
expansion free cases. Jawad \& Rani (2015) examined the instability
ranges taking into account shear-free condition for Newtonian and
post-Newtonian regimes via adiabatic index. These works are only
discussed for anisotropic fluid. In order to compare the results of
present paper, we analyze that results depend on the physical
quantities, like energy density, pressure, curvature terms and mass
of the cylinder. However, to make a correspondence with the results
of isotropic sphere, we have established possibilities (after
Eq.(48) as I, II, III) on these physical quantities in each case
(isotropic as well as anisotropic fluids) to have numerical results
like Chandrasekhar (1964).

So far we know that the cosmographic features of $f(T)$ gravity
mimics LCDM model as well as phantom dark energy models. Although
there are some ambiguities about the validity of solar system tests
for $f(T)$ due to the absence of the Schwarzschild solution as the
vacuum (Rodrigues et al. 2013), but the dynamics of stellar objects
and what we studied here as the (in)stability of cylindrical objects
provides a good sample to check the validity of $f(T)$ gravity in
the astrophysics. Briefly the stability conditions were obtained in
our paper have significant physical meaning in comparison to the
classical results. Furthermore, the model which we studied here, the
torsion based version of the Starobinsky model tested among several
types of models with cosmological data, so we believe that our paper
will be useful for astrophysical tests of compact objects in $f(T)$
gravity.

\end{document}